\documentclass[preprints,article,accept,oneauthor,pdftex]{Definitions/mdpi} 

\firstpage{1} 
\pubvolume{1}
\issuenum{1}
\articlenumber{0}
\pubyear{2021}
\copyrightyear{2020}
\datereceived{} 
\dateaccepted{} 
\datepublished{} 
\hreflink{https://doi.org/} 

\Title{Solar and anthropogenic climate drivers: an updated regression model and refined forecast}

\TitleCitation{Solar and anthropogenic climate drivers: an updated regression model and refined forecast}

\Author{Frank Stefani $^{1}$ \orcidA{}}

\AuthorCitation{Stefani, F.}

\address{%
$^{1}$ \quad Helmholtz-Zentrum Dresden-Rossendorf, 
Institute of Fluid Dynamics,
Bautzner Landstr. 400, 01328 
Dresden, Germany; F.Stefani@hzdr.de}

\corres{Correspondence: F.Stefani@hzdr.de}

\abstract{In a recent paper \cite{Stefani2021}, attempts were made to 
quantify the respective solar and anthropogenic 
influences on the terrestrial climate,
and to cautiously predict the global mean temperature 
over the next 130 years. 
In a double regression analysis, both 
the binary logarithm of carbon dioxide concentration 
and the geomagnetic aa-index were used as 
predictors of the sea surface temperature (SST) 
since the mid-19$^{\rm th}$ century. 
The regression results turned out to be
sensitive to end effects, leading
to a disconcertingly broad range of the climate 
sensitivity between
0.6\,K and 1.6\,K per doubling of CO$_2$ 
when varying 
the final year of the data used. 
The aim of this paper is to 
significantly narrow down this range. To this end, the correlations between the two predictors and the dependent variable (SST) are analysed in detail. It is demonstrated that the SST can be predicted until around 2000 almost perfectly using only the aa-index, whereas for later periods the role of CO$_2$ increases significantly.
Therefore, the weight of the aa-index is fixed
to its very robust outcome (around 0.04 K/nT) from the 
single and double 
regressions up to 1990. The SST data, reduced by the aa-contribution thus
specified, are then
used in a single regression with 
CO$_2$ as the only remaining predictor.
This results in a significant reduction in the range of CO$_2$ sensitivity, narrowing it to 1.1–1.4\,K.
Given the exceptionally high temperatures in recent 
years, these values are considered a kind of upper 
limit that could still be subject to downward corrections when future data are incorporated.
Based on this estimate, the
temperature forecast until 2100 is refined by using more precise 
predictions of the aa-index and the paths of atmospheric CO$_2$ content which are based on constant emission scenarios combined with a linear sink model. With the exception of the most ``pessimistic'' variant, the  temperature is predicted to remain below the extraordinarily high value 
measured in 2024.
}


\begin{document}

\section{Introduction}

The true value of climate sensitivity - the global mean temperature increase for a doubling of CO$_2$ content in the atmosphere - is of the utmost importance in determining the urgency (or serenity) with which future emissions of this greenhouse gas should be addressed.
 Yet, the broad range of its most
prominent metric - Equilibrium Climate Sensitivity (ECS) - has barely narrowed since the early times of the Charney report \cite{Charney1979}. 
The replacement of the 1.5-4.5\,K range,
favored throughout the first five IPCC reports, 
with the new value 2.0-5.0\,K \cite{AR6} 
can hardly be considered a progress in this respect.
For the ``very likely" 1.2-2.4\,K range of the Transient Climate Response (TCR),
which is widely considered to be 
``more relevant for 
predicting climate change over the next century'' 
\cite{Knutti2017}, the situation is only slightly better.

One of the most likely reasons for these estimates not improving is the systematic underestimation of the climatic impact of variations in solar activity. 
This negligence is surprising, given that the significant influence of these variations has long been a well-accepted fact in climate science.
A case in point is the solar triggering of Bond events throughout the Holocene, as evidenced by the close correlation between changes in the production rates of cosmogenic nuclides and centennial-to-millennial-scale changes in drift ice proxies \cite{Bond2001}. More short-term temperature variations, and their link to solar activity changes since the beginning of the 
telescope time, were evidenced by \cite{Lean1995}.
In view of the partly overlapping authorship,
it is entertaining to observe how the
clear acknowledgment of solar influence in 
that work gave way to the claimed dominance of 
CO$_2$ in the 20th century, when the iconic hockey 
stick graph \cite{Mann1998} was published three years 
later\footnote{While the mathematical issue of ``short centering'' behind the generation of hockey-stick-shaped curves was thoroughly discussed in \cite{McIntyre2005}, less attention was paid to how the steep temperature increase of 0.5\,K between 1900 and 1940 could be plausibly explained by a corresponding increase in the binary logarithm of CO$_2$ of only 0.05.
Since this would require a TCR of 10\,K, which may seem excessive even to those who advocate high climate sensitivity, it is legitimate to consider other causes besides anthropogenic ones.
Interestingly, these very points had already been addressed in 1938 during the panel debate that followed Callendar's presentation of his seminal paper \cite{Callendar1938}.}. 

One of the issues in understanding the role of the Sun is the purely known variability of total solar irradiance (TSI) over decadal and centennial timescales.
This problem was recently spotlighted by Connolly et al. \cite{Connolly2021,Connolly2023}
who assessed TSI models with both very 
low (appr. 1\,W/m$^2$, e.g. \cite{Steinhilber2009}) and very high (appr. 6\,W/m$^2$, e.g \cite{Egorova2018}) variations since the Maunder minimum. They ultimately concluded that 
``it is still unclear whether the
observed warming is mostly human-caused, mostly natural or some combination of both'' \cite{Connolly2023}.
Evidently, this assessment contradicts the 
more definite claims of \cite{Yeo2020} 
and \cite{Penza2024}
that the TSI level of the Sun 
during grand minima was only 
around
2\,W/m$^2$ below its more recent 
maximum level.

As interesting as this TSI-variability issue may be, 
it masks the deeper problem of whether the solar 
influence on climate predominantly occurs through 
a bottom-up or top-down mechanism.
\cite{Gray2010,Georgieva2012,Georgieva2023}. 
Indeed, the focus on TSI variation may be rather
misleading if one of the top-down mechanisms were 
to turn out to be dominant. Among those, the 
following candidates have been discussed in the 
literature:
the effect of the modulation of cosmic rays by the 
solar magnetic field on aerosols and clouds \citep{Svensmark1997,Soon2000b,Shaviv2003,Svensmark2017,Svensmark2021};
downward winds following geomagnetic storms in the 
polar caps of the
thermosphere, penetrating the stratosphere and the 
troposphere \citep{Bucha1998};  
the impact of the solar wind on global electric currents \citep{Tinsley2000,Tinsley2008}; absorption of the strongly varying solar UV radiation by stratospheric ozone, and subsequent changes of large-scale circulation patterns such as the the stratospheric polar vortex or the tropospheric North Atlantic Oscillation
\citep{Labitzke1988,Haigh1994,Soon2000,Georgieva2012,Silverman2018,Veretenenko2020,Georgieva2023}.

The latter variant looks particularly 
interesting in view of the remarkable 
agreement between the spectral peaks 
obtained from a 8500\,year sediment stack from 
Lake Lisan \cite{Prasad2004} with those
provided by our synchronized solar dynamo model
\cite{Stefani2024}.
This nexus indeed points to a 
significant solar effect on 
atmospheric circulation pattern in general, and on the paths of North Atlantic cyclons in particular, as recently discussed by \cite{Veretenenko2023,Veretenenko2024}.
Such mechanisms are, of course, much more complicated (potentially even involving reversals in the correlations between solar activity and atmospheric parameters, see \cite{Georgieva2023}) than a pure bottom-up effect resulting from TSI variations 
only\footnote{What also remains to be understood in more detail is the link between the observed widening of
the Hadley circulation (and the related poleward shift in midlatitude storm tracks) with the change of the 
cloud areas representing, respectively, 
the midlatitude and tropical storm zones. As demonstrated by 
Tselioudis et al. \cite{Tselioudis2025}, this area change constitutes the largest contribution to the positive solar absorption trend that may provide ``a crucial missing piece in the puzzle of the 21$^{\rm st}$ century increase of the Earth's solar absorption''.
For a discussion of the clouds' dominant role for the declining outgoing shortwave radiation as the most important contributor for a positive net flux at the top of the atmosphere, see also \cite{Dubal2021}.}.

In view of those ambiguities, Scafetta \cite{Scafetta2023a} proposed to use the TSI just as possible proxy of a more general {\it total solar activity} (TSA) whose climatic impact
might well be much stronger than what could be expected from the TSI alone. He found a total solar impact on climate change that is 4–7 times greater than its TSI effect, along with a significantly reduced effect of CO$_2$. The  ECS-range found by him was 0.8–1.4\,K, with a mean of around 1.1\,K. Such low values were also plausibilized in
another paper showing that only the low-ECS Global 
Circulation Models (GCM) can be considered consistent with the surface-based global temperature \cite{Scafetta2023b}. 

With the TSI being deprived of its priority role, I have recently considered the role of the 
geomagnetic aa-index in relation to climate
\cite{Stefani2021}. While inspired by previous work \cite{Cliver1998,Mufti2011}, this index was not chosen
as it fits to certain temperature 
data\footnote{Such a claim was made in \cite{Chatzi2025}: ``Stefani (2021) explicitly chose to associate the geomagnetic aa-index with global sea surface temperatures not because that is the most logical predictor between the Sun and Earth’s climate, but because it presents a strong match between a solar record and certain Earth temperature records''. Fortunately, improper insinuations of this kind are almost unique to the non-peer-reviewed literature.}, but because it provides -- in contrast to the TSI -- an unambiguous  proxy of solar activity that has been continuously measured since 1844 (for details concerning a minor correction after a station change in 1957, see \cite{Lockwood2014}).

In a double regression analysis this aa-index and 
the binary logarithm of carbon dioxide concentration were used 
as predictors for the sea surface temperature (SST) 
since the middle of the 19$^{\rm th}$ century. 
Alas, the regression results turned out to be
sensitive to end effects, leading
to a disconcertingly broad range of the climate 
sensitivity 
(of the TCR type) between
0.6\,K and 1.6\,K per doubling of CO$_2$ 
when varying 
the end year of the utilized data.

 The aim of this paper is to refine the findings of \cite{Stefani2021} by incorporating new data from 2019 to 2024, while retaining the core methodology. In particular, it avoids taking into account additional predictors, such as the influence of volcanoes and the El Ni\~{n}o–Southern Oscillation (ENSO). Although including the latest 
El Ni\~{n}os might help to 
mitigate the end effects of the regression, it has also been noted \cite{Schmidt2024} that the unexpectedly steep temperature increase in 2023 was hardly be explainable by ENSO effects, let alone the slowly increasing CO$_2$.
Although the Hunga volcano's injection of sulphur aerosols and water vapour into the stratosphere is certainly one possible ``culprit'' of the sudden warming{\footnote{While the expected time-ordering of a cooling effect of aerosols through the first months after eruption and the warming effect of water vapour in the following years \cite{Zhuo2025} would fit well to the steep temperature increase in 2023-24, and the slow decline in 2025 \cite{Vinos2025}, a detailed understanding of the role of stratospheric water vapor in modulating long-term atmospheric chemistry and dynamics is still pending \cite{Bednarz2025,APARC2025}.}}, I
remain agnostic as to its ultimate cause and simply acknowledge that current models may not accurately reflect all temperature variations over periods of a few years.

How can this problem be coped with when staying in the two-predictor framework?  The approach adopted here takes the significant change in the relative shares of the two predictors over time into serious consideration. As will be demonstrated, until the end of the 20$^{\rm th}$ century the aa-index alone is essentially sufficient to describe the temperature evolution.  In fact, the adjusted $R^2$ value of the single regression including only aa will be shown to be higher than that of the double regression including both aa and CO$_2$. Since the climate sensitivity on aa, arising both from the double and the single egression until the end of the 20$^{\rm th}$ century, is quite stable over some decades, this value (appr. 0.04 K/nT) will be considered valid for the later decades as well.

Based on this reasoning, I deliberately commit the ``statistical sin'' of replacing the  ``correct'' double regression with a single regression of the temperature data, which has been reduced by the physically reasonable aa-contribution beforehand, on CO$_2$. This approach prevents the exceptionally high temperatures of recent years from completely ruining the resulting aa-sensitivity by forcing it to very low values in the double regression.
This way, the climate-sensitivity on CO$_2$ is narrowed to a range of 1.1-1.4\,K for the modified single regression, as opposed to 1.75\,K for the double regression.

The second part of the paper then provides a provisional forecast of the global mean temperature up to the end of the century. Once again, this will rely on the relevant section of \cite{Stefani2021}, in which  the future of the aa-index was predicted on the basis of the aforementioned solar dynamo synchronisation model, while considering different CO$_2$ scenarios.
However, some new aspects that have emerged since the publication of \cite{Stefani2021} will be incorporated here.  The first one, related to the aa-index, relies on the improved trustworthiness of our synchronised solar dynamo model, which by now reproduces both the dominant periods of the Lake Lisan sediment data \cite{Prasad2004,Stefani2024} and the dominant period of the solar Quasi Biennial Oscillation (QBO) \cite{Stefani2025} with remarkable accuracy.
Therefore, unlike in \cite{Stefani2021}, the dominant periods will be  fixed beforehand to 193, 90 and 57\,years when forecasting aa.

Guided by the recent IEA report \cite{IEA2025}, I will consider constant annual CO$_2$ emissions of 30, 40 and 50\,Gt, and combine these with a simple linear sink model, the plausibility of which has recently been bolstered by the work of Dengler \cite{Dengler2023,Dengler2024}.

Of the nine models thus assessed, the most plausible one predicts a temperature increase of around 0.6\,K by the end of the century compared to the 1961–1990 reference period. 
The spread around this value, resulting from varying the climate sensitivities and the emission scenarios, amounts to 
appr. $\pm 0.3$\,K. 

The paper will end with a summary of the results and some conclusions.

\section{Data base and methodological restrictions}

As in \cite{Stefani2021}, I  perform  a 
double regression of the sea surface temperature data (dependent variable) on the geomagnetic aa-index 
\cite{Mayaud1972} and the binary logarithm of CO$_2$ (independent variables). This double regression is similar in spirit to that of \cite{Soon1996} who used, however, the length of the sunspot
cycle and the mean sunspot number as proxies
of solar activity.
Clearly, including volcanic and ENSO forcings, as in references \cite{Lean2008} and \cite{Scafetta2023a}, could  improve the study, especially towards the end of the time period, when El Ni\~{n}o events have a significant impact.
Furthermore, the exceptionally steep temperature increase at the end of our database, which may have been caused by the Hunga explosion, might require special consideration.
In general, however, the double regression is considered well-justified by the use of moving averages (usually 11 and 23 years), which largely eliminate short-term contributions.

I also refrain from studying possible time lags between the dependent and independent variables, which were shortly discussed in \cite{Stefani2021} and in more detail  by \cite{Scafetta2023a}.

With all such possible improvements being postponed to future work, the data-sets used are essentially the same as in \cite{Stefani2021}, but now with 2024 instead of 2018 as the final year.  The starting point is always 1850, except for a later multiple-period fit of the aa-index, for which 1844 will be used as starting point. 

As for the temperature, the Hadley Centre Sea Surface Temperature (HadSST) data are used in its updated version HadSST.4.2.0.0, available from 
www.metoffice.gov.uk, which give the sea surface temperature anomaly from 1850 until 2024, relative to the 1961-1990 average. The data are shown as open circles in Figure 1a, alongside two exemplary centred moving averages with moving average windows (MAWs) of 11\,years (full line) and 23\,years (dashed line). These curves exhibit a typical temporal structure: a slow decline between 1850 and 1905, a sharp rise between 1905 and 1940, a gradual decline until 1970 followed by a sharp increase until 1998.  Looking at the last decades, the first thing to notice is the ``hiatus'' between 1999 and 2014, followed by some strong El Ni\~no events and finally the sharp increase to almost 1\,K in 2023-2024.

The bulk of the aa-index data, between 1868-2010, was obtained from the ftp-server ftp.ngdc.noaa.gov.
As in \cite{Mufti2011}, the early segment between 1844 and 1867 was taken from Nevanlinna and Kataja 
\cite{Nevanlinna1993}. The latest segment,
between 2011 and 2024, was obtained from
www.geomag.bgs.ac.uk.
All of the aa-index data were averaged annually to provide a homogeneous dataset. Figure 1b shows the annual aa-index data together with their 11-year and 23-year moving averages. Even by visual inspection, we can see a remarkable similarity in shape between the temperature and the aa-index between 1850 and 1990. After 1995, however, the aa-index declines while the temperature continues to increase. 

The CO$_2$ data, the binary logarithm of which is shown in Figure 1c, were obtained from www.co2.earth,  www.co2.earth  and gml.noaa.gov.

\begin{figure}[H]		
\includegraphics[width=0.7\linewidth]{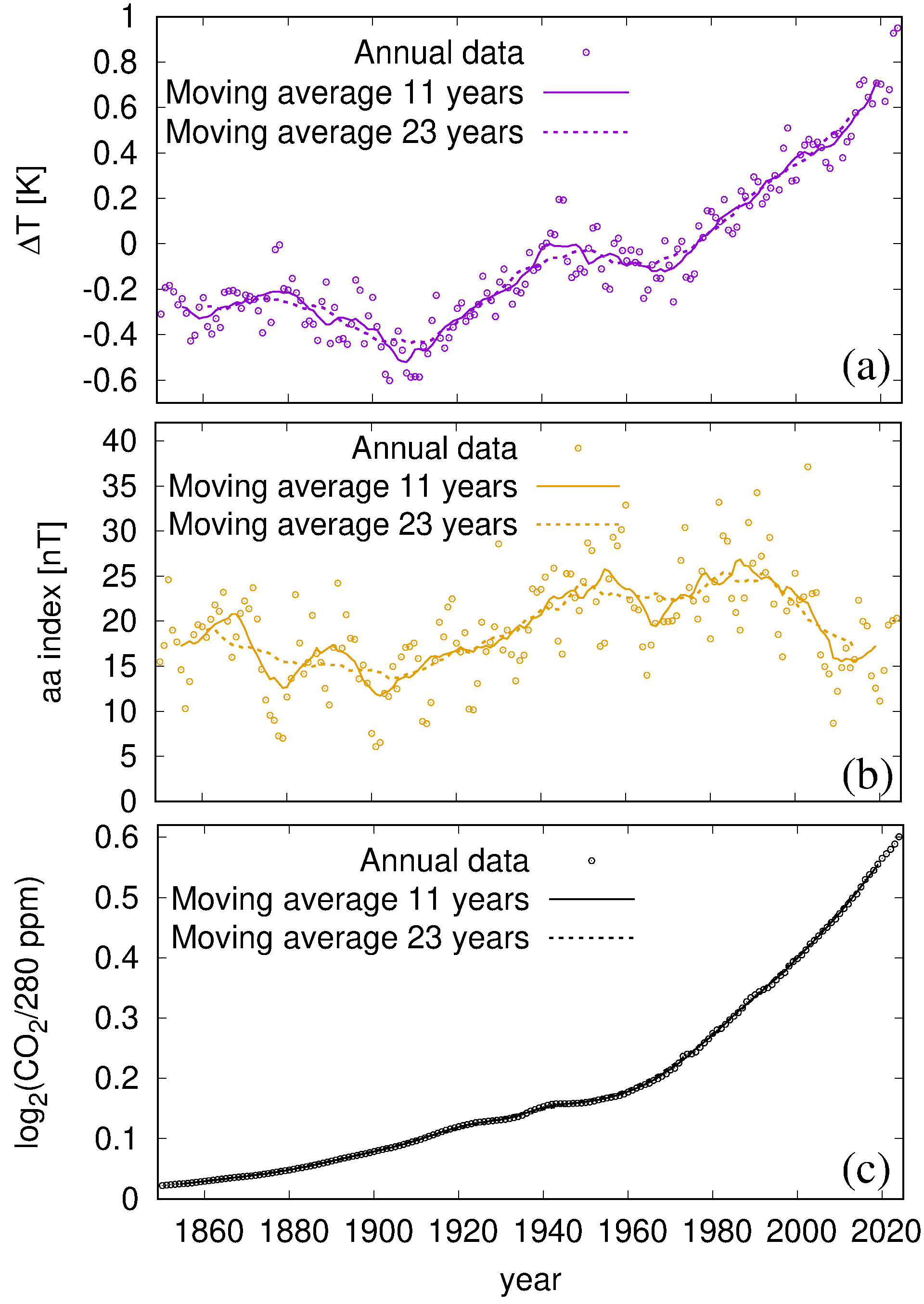}
\caption{Data on the HadSST sea surface temperature anomaly $\Delta T$ (a),
the aa-index (b), and $\log_2$ of the ratio of the CO$_2$ concentration to the reference value 
of 280 ppm (c). Centred moving averages with windows of 11 years (full lines) and 23 years (dashed lines) complement the annual data between 1850 and 2024.
The sources of the data are described in the text.}
\label{FIG:1}
\end{figure}

\section{One double and two single regressions}

As in \cite{Stefani2021} I will model the temperature
(dependent variable) by using the ansatz
\begin{equation*}
\Delta T^{\rm model}=w_{\rm aa} \cdot {\rm aa} + w_{\rm{CO_2}} \cdot \log_2({\rm CO_2}/280\, {\rm ppm})
\end{equation*}
with the respective weights $w_{\rm aa}$ and $w_{\rm{CO_2}}$ 
for the two independent variables $\rm aa$ and CO$_2$,
and compare them with the measured data $\Delta T^{\rm meas}$.

In \cite{Stefani2021} it was shown that the
outcome of the double regression is contingent on the final year of the used data. As for $w_{\rm{CO_2}}$, a value of 1.6\,K was obtained when using 2018 as the end year, 
and 0.6\,K when using 2008.
The corresponding sensitivities $w_{\rm aa}$ were 
0.017\,K/nT and 0.032\,K/nT, respectively.

 Apart from potentially unconsidered factors, such as ENSO or the Hunga eruption, this sensitivity to end effects is also due to a high degree of collinearity between the two independent variables, aa and CO$_2$. This means that quite different combinations of the two weights can potentially produce very similar reconstruction qualities.  In Figures 4, 5 and 6 of \cite{Stefani2021} this co-linearity was visible
in the downward-directed ellipses that appear when plotting the  fraction of variance unexplained (FVU) 
in the plane spanned by $w_{\rm aa}$ and $w_{\rm{CO_2}}$.
Along the long axis of those ellipses the FVU 
changes only weakly.

The strong end effect in this picture can be explained as follows: Using a year too early means that CO$_2$ did not have enough time to be reflected in the temperature curve, resulting in an underestimation of $w_{\rm{CO_2}}$.
By contrast, for late end years, the weight may shift unduly from aa to CO$_2$, particularly if those final years are particularly hot, for example due to El Ni\~{n}o conditions or other factors.

In the following I will try to disentangle this problem in a systematic way. I start with the standard ``statistically correct'', double regression, which is computed for all end years between 1950 and 2024. 
In parallel, two single regressions are carried out, with either aa or CO$_2$ as the sole independent variable. 
I commit this ``statistical sin'' very consciously
(being fully aware of the criticism of 
\cite{Connolly2021} by \cite{Richardson2022}), to see how the results, and their 
$R^2$ values, compare 
with those of the ``correct'' double regression.
This comparison will reveal some interesting insights into the dominance of each factor and how it changes over time.

Let us start with an MAW of 11 years which mainly 
``ìrons out'' out the effect of the Schwabe cycle 
of the solar dynamo on aa. 
Figure 2a shows the $R^2$ value for the
double regression and the two single regressions 
on either aa or CO$_2$. Interestingly, until 1990 the two $R^2$ values for the double regression and the single one on
aa are nearly identical, reaching values around 0.7 towards this year. Meanwhile, $R^2$  for the single regression on CO$_2$ is significantly smaller.
Only from the year 2005 onward, the  
single-regression $R^2$ for CO$_2$ exceeds that for aa, and approaches that for the double regression in the 2020s. 
For those late years, we observe a drastic reduction of  $R^2$ for the single regression on aa to a very low value of 0.2.

Figure 2b shows the adjusted variant of $R^2$, coined
${\overline{R}}^2$, a useful indicator of a model's true predictive power, which penalizes the addition of irrelevant variables and only increases if a new predictor significantly improves the model's fit.
In our special case, it can be computed as ${\overline{R}}^2=1-{\rm FVU} \times (N_{\rm y}/{\rm MAW}-1)/(N_{y}/{\rm MAW}-1-p)$, where $N_{\rm y}$ is
number of considered years and $p$ the number of 
explanatory terms (i.e., 2 for the double and 1 
for the single regression).

 While its shape is not significantly different to that of the $R^2$ curve in Figure 2a, it is interesting to note that until 1985 ${\overline{R}}^2$ for the single regression on aa is even 
{\it higher} than that for the double regression.  Therefore, it is fair to say that, until this moment, the goodness of fitting $\Delta T$ by aa alone is higher than fitting it by aa and CO$_2$ jointly.

Figure 2c shows that the resulting weight of aa until this moment is relatively constant (around 0.03\,K/nT), with only a minor difference between the double and the single regression. Later, however, the value from the double regression drops to 0.01\,K/nT, while the value from the single regression is no longer meaningful in view of the drastically falling $R^2$ value. In contrast, the double regression value for $w_{\rm{CO_2}}$ remains low at 0.5 K until 1990, after which it increases steeply to reach 1.8 K by the year 2024.

\begin{figure}[H]		
\includegraphics[width=0.70\linewidth]{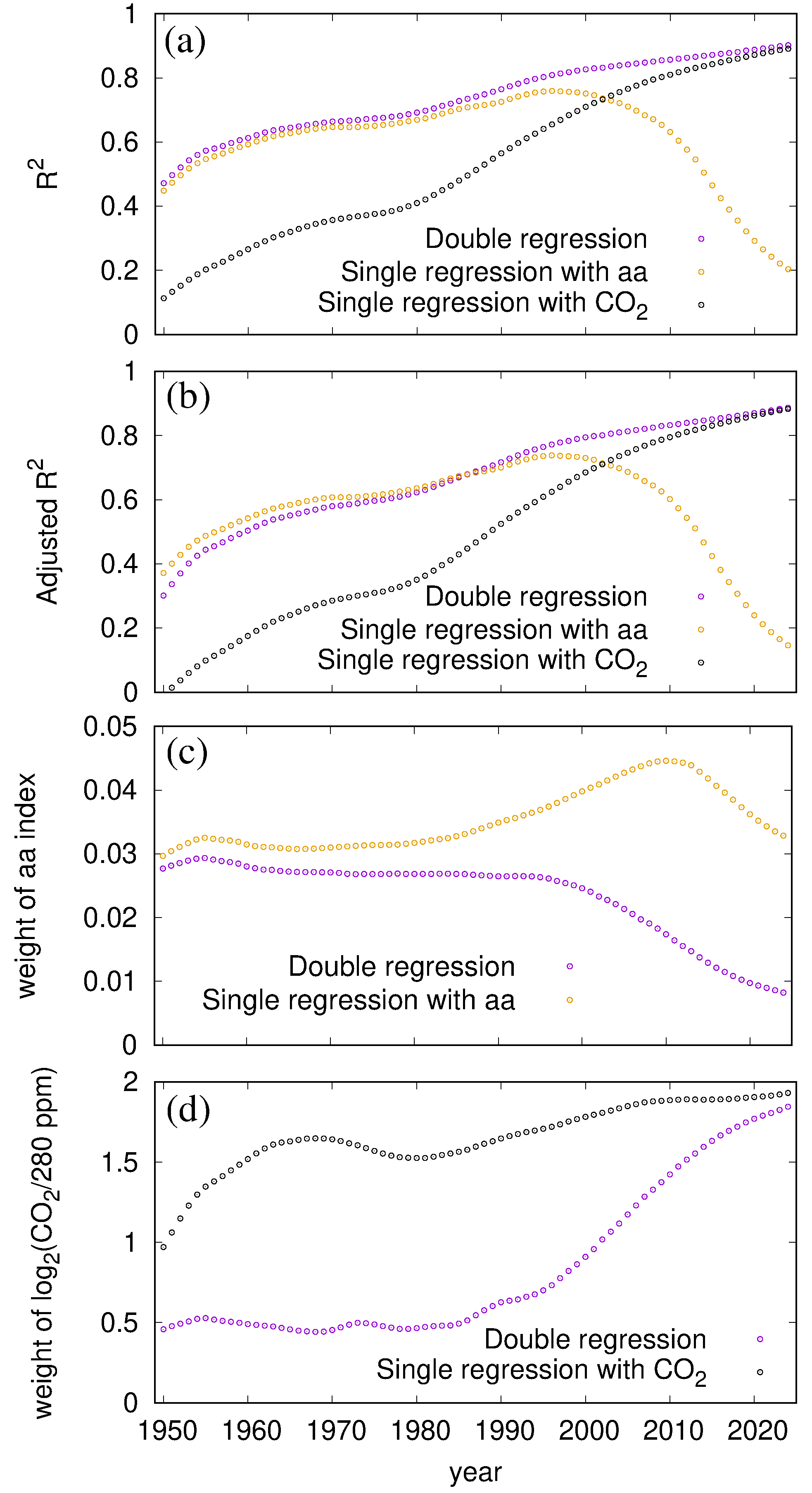}
\caption{Regression results in dependence on the chosen end year for a moving
average window (${\rm MAW}$) of 11 years. (a)  $R^2$ for the double regression and the two single regressions 
with either the aa index or the binary logarithm of CO$_2$ as the independent variable. (b) The same as (a), but 
for the adjusted version ${\overline{R}}^2$.
(c) Resulting $w_{\rm aa}$ for the double regression and the single regression
with aa as the only independent variable. (d) 
Resulting $w_{\rm CO_2}$ for the double regression and the single
regression with CO$_2$ as the only independent variable.}
\label{FIG:fig2}
\end{figure}

Figure 3 shows the qualitatively similar behaviour when an MAW of 23 years is chosen.
Here, ${\overline{R}}^2$ for the single regression on aa remains higher than that for the double regression even until the year 2005.
Again one could argue that, until this year, including CO$_2$ would only worsen the goodness of fit.
For this MAW, the long-term result for w$_{\rm aa}$ is around 0.04\,K/nT, while it drops to 0.01\,K/nT for the final year 2024.
Accordingly, w$_{\rm CO_2}$ remains close to zero until 2000, after which it increases steeply to reach 1.7 by 2024.

\begin{figure}[H]		
\includegraphics[width=0.70\linewidth]{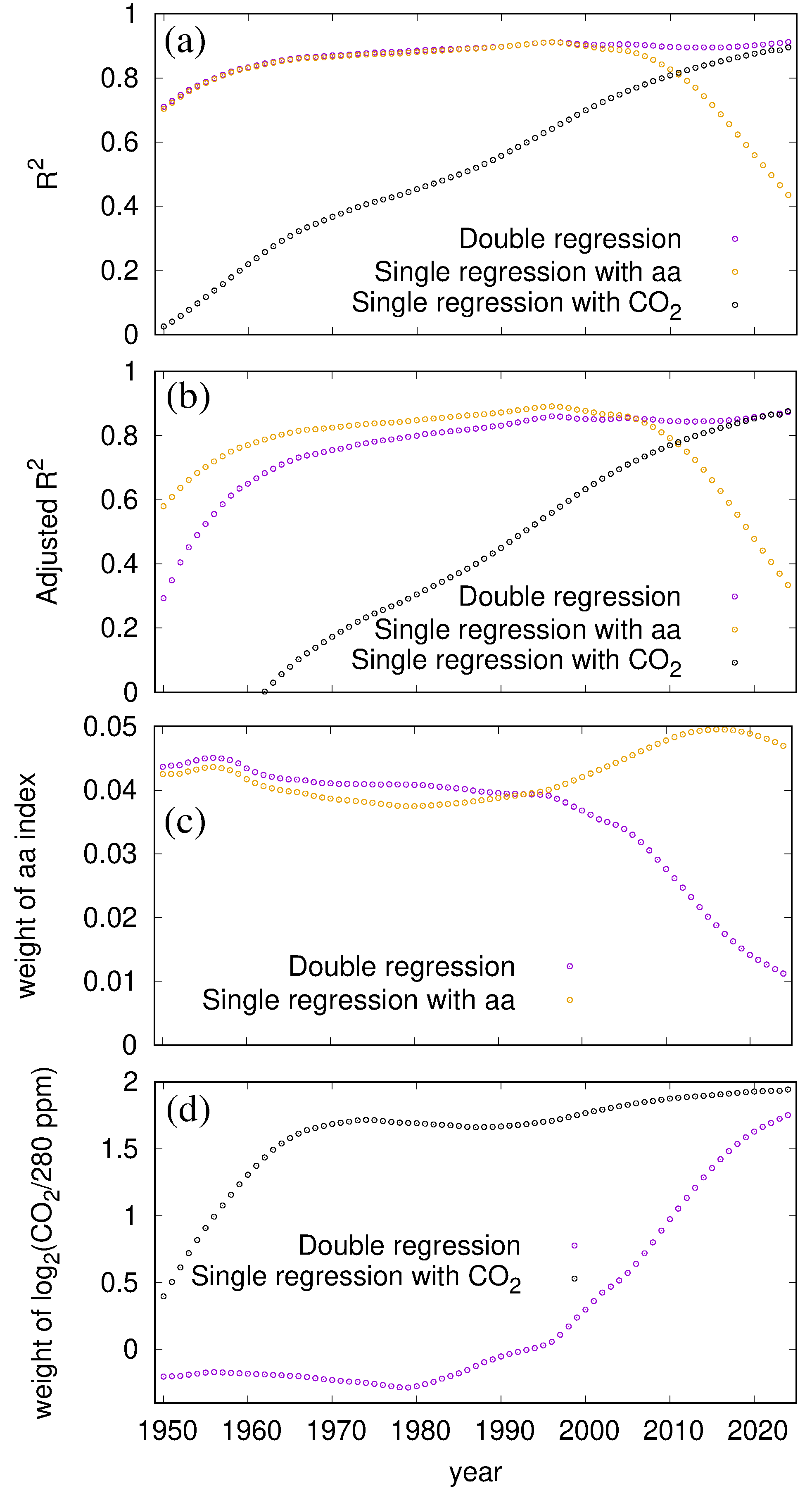}
\caption{The same as Figure 2, but with an MAW of 23 years.
Note that in (b)  the adjusted version ${\overline{R}}^2$ formally 
acquires, for early years, 
negative values that are not shown.}
\label{FIG:fig3}
\end{figure}

Clearly, there is a problem here. On one hand, until the end of the 20$^{\rm th}$ century, $\Delta T$ turns out to be more or less perfectly predicted by aa alone, resulting in a quite stable aa-sensitivity of appr. 0.04\,K/nT (if we chose an MAW of 23 years for which the ${\overline{R}}^2$ values of appr. 0.9 are higher than those for 11 years).
On the other hand, the steep temperature increase over the past decade requires not only an increasing share of CO$_2$ (which is well understandable in view of its steep increase, see Figure 1c), but also a dramatic {\it decrease} of $w_{\rm aa}$ which is much harder to explain.
In our view, the only explanation is that some part of the $\Delta T$-increase over the last years is due to some other effects (beyond CO$_2$), like ENSO or the  Hunga volcona, 
which are not properly included in our 
simple model.

Therefore, we must either accept an unreasonably large reduction in the aa-sensitivity over the last two decades (compared with its long-term stable average), or attribute some of the recent steep warming to factors other than CO$_2$ or solar activity.

 The second approach will be tentatively pursued in the following (its correctness 
 will ultimately be judged by the temperature development over the next years). Specifically, the long-term result of about 0.03-0.05\,K/nT for $w_{\rm aa}$ will also be used for the 
later times, for which the double regression would predict a much reduced value. Having thus fixed $w_{\rm aa}$, one is left with another single regression on CO$_2$, but this time of a modified $\Delta T$ that is reduced by the 
aa-contribution weighted by the  $w_{\rm aa}$ derived beforehand.

This procedure is illustrated in Figure 4 (for an MAW of 11 years) and Figure 5 (for an MAW of 23 years). Here the violet curves for $R^2$ (a) and $w_{\rm{CO_2}}$ (b) correspond, respectively, to those from  panels (a) and (d) of Figures 2 and 3.
The black curves, however, result when first subtracting from $\Delta T$ an aa-contribution weighted by a fixed $w_{\rm aa}$. For the latter we chose a certain variety, ranging from 0.03\,K/nT (according to Figure 2c), via 0.04\,K/nT (according to Figure  3c) to 0.05\,K/nT (
the uppermost value in Figure 3c).

\begin{figure}[H]		
\includegraphics[width=0.70\linewidth]{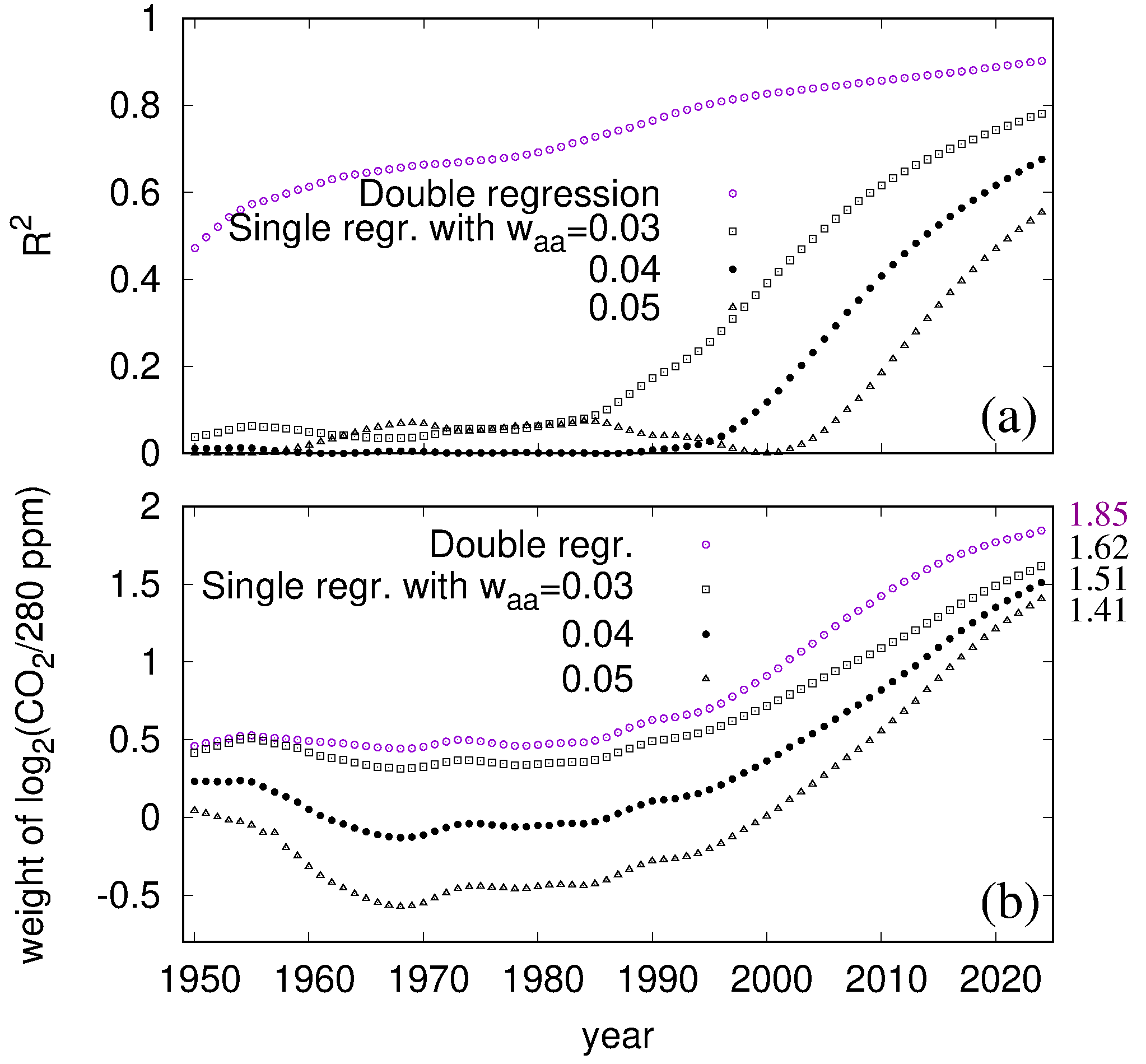}
\caption{Regression results in dependence on the chosen end year for a moving
average window (${\rm MAW}$) of 11 years. (a)  $R^2$ for the double regression (as in Figure 2a) and the three single regressions on the binary logarithm of CO$_2$ as the independent variable, when subtracting beforehand from $\Delta T$ the aa-contribution weighted with 
three different values of $w_{\rm aa}$ (0.03\,K/nT, 0.04\,K/nT, and 0.05\,K/nT).
(b) Resulting $w_{\rm CO_2}$ for the double regression and the three single
regressions.}
\label{FIG:fig4}
\end{figure}

\begin{figure}[H]		
\includegraphics[width=0.70\linewidth]{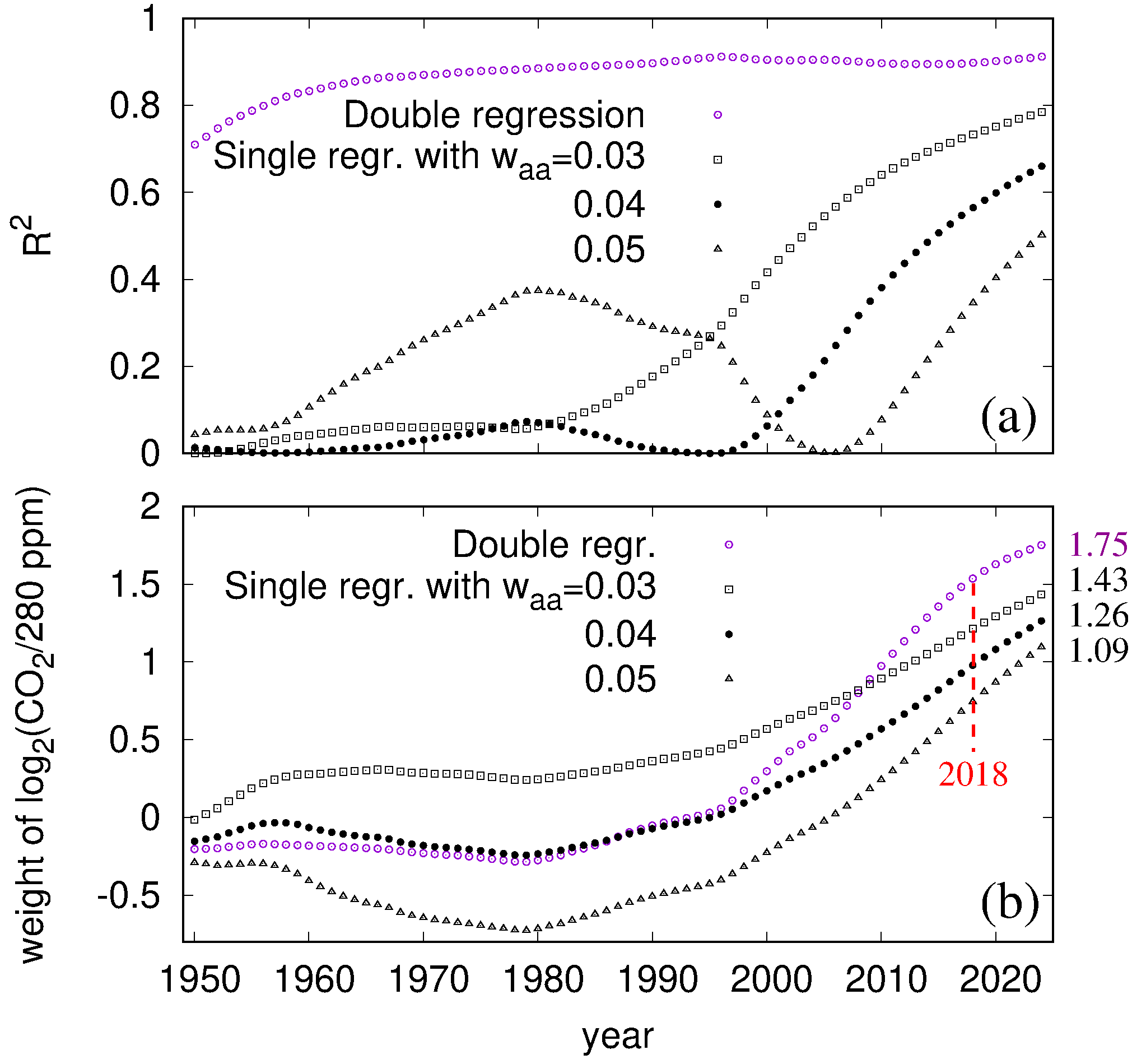}
\caption{Same as Figure 4, but for an MAW of 23 years.}
\label{FIG:fig5}
\end{figure}

Note that constraining $w_{\rm aa}$ beforehand has a 
non-trivial impact on the subsequent single regression on CO$_2$.
On the one hand it tends to require {\it less}
of $w_{\rm{CO_2}}$ since a larger proportion of the increase between 
1850 and 2000 has  already been 
attributed to aa. On the other hand, for the late interval, starting from 1990, even {\it more} $w_{\rm{CO_2}}$ is required to compensate for the strongly negative aa-contribution due to the steep aa-decline during those years.
Obviously, the combined effect of these two tendencies leads to a moderate reduction in the resulting $w_{\rm{CO_2}}$, compared to the double regression.

The $w_{\rm{CO_2}}$ values resulting in the final year, 2024, are indicated on the right-hand ordinate axis of Figures 4b (for an MAW of 11 years) and 5b (for an MAW of 23 years).
The latter values, 1.09-1.44\,K are situated in the upper half of the 0.6-1.6\,K range that had been predicted in \cite{Stefani2021}.  This upward trend is mostly due to the high temperatures that have been recorded over the past six years.
Had we carried out the same procedure based on the previous dataset (up to 2018), we would have obtained a range of 0.8–1.2\,K (see the red line in Figure 5b), which lies slightly below the middle of the 0.6–1.6\,K interval.

Figure 6 focuses on the MAW of 23 years and illustrates how the resulting combinations of $w_{\rm{CO_2}}$ and $w_{\rm{aa}}$ play out in comparison with the original $\Delta T$ data. 
Figure 6a shows a reasonable behaviour in recent years when the results of the ``correct'' double regression are chosen, i.e., $w_{\rm{aa}}=0.011$\,K/nT 
and $w_{\rm{CO_2}}=1.75$\,K). However, this comes at the cost of barely reflecting the temperature variability between 1850 and the end of the 20th century.
This variability is better reflected 
when accepting for $w_{\rm{aa}}$ the long-term outcomes of 0.03\,K/nT (b), 0.04\,K/nT (c) or 0.05\,K/nT (d), together with the corresponding 
$w_{\rm{CO_2}}$ values of 1.43\,K,
1.26\,K and 1.09\,K.
However, the temperature of the latest years, as might be expected, is becoming less and less accurately represented.

\begin{figure}[H]		
\includegraphics[width=0.70\linewidth]{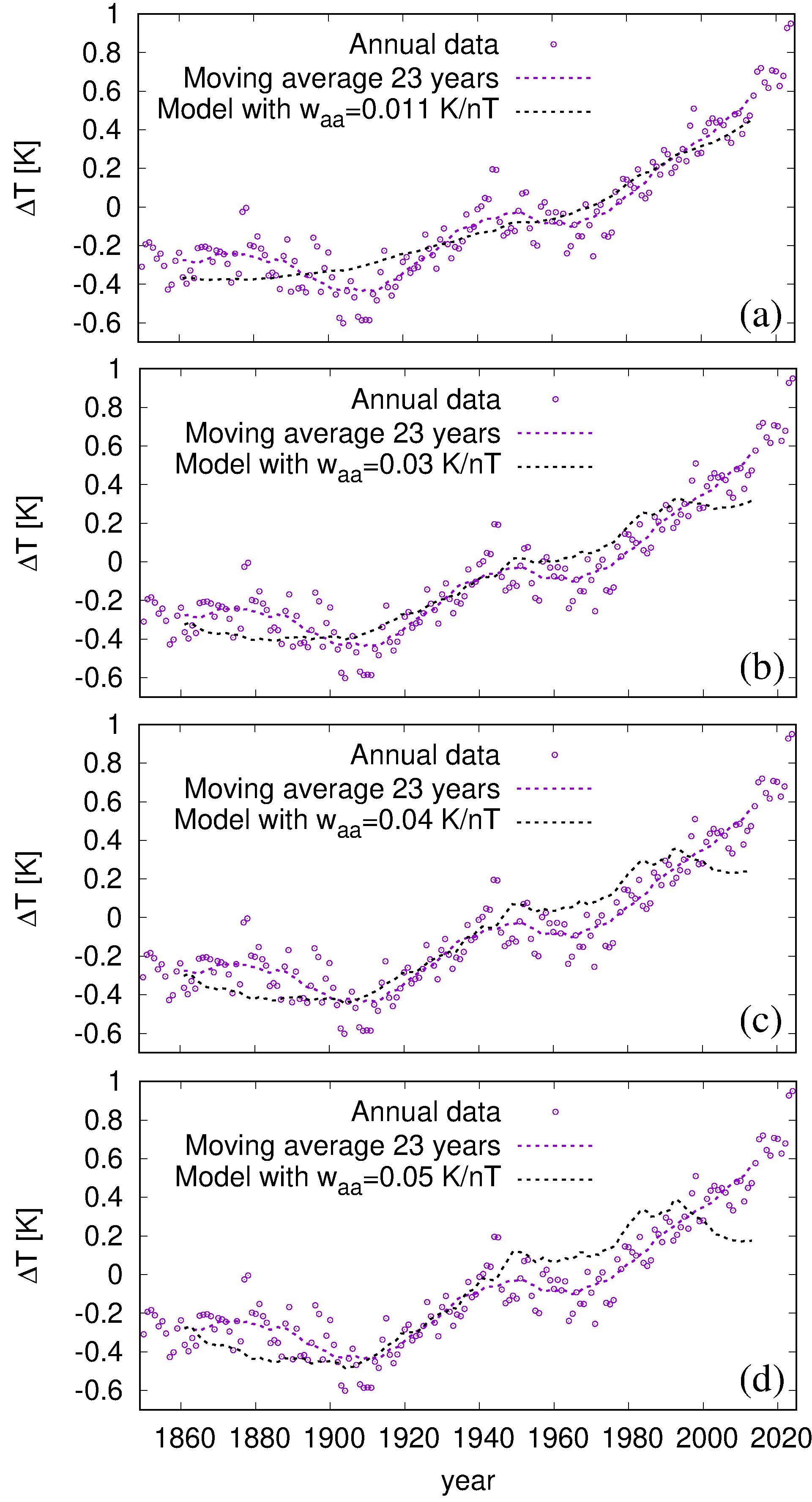}
\caption{Data of $\Delta T$ and their average with an MAW 23 years as in Figure 1a, together with reconstructions with different sensitivity parameter combinations. (a) 
$w_{\rm{aa}}=0.011$\,K/nT 
and $w_{\rm{CO_2}}=0.175$\,K/nT, as resulting from the double regression until the end year 2024.
(b) $w_{\rm{aa}}=0.03$\,K/nT 
and $w_{\rm{CO_2}}=1.43$\,K. (c) $w_{\rm{aa}}=0.04$\,K/nT 
and $w_{\rm{CO_2}}=1.26$\,K. (d)
$w_{\rm{aa}}=0.05$\,K/nT 
and $w_{\rm{CO_2}}=1.09$\,K. }
\label{FIG:fig6}
\end{figure}

\section{Predictions}

In the preceding section, regression analyses of data from the past 175 years were used to derive a range of plausible combinations of the respective sensitivities on the aa-index and the binary logarithm of CO$_2$. 
In the following, I will employ these results 
to forecast the 
global mean temperature until 2100.
While forecasts up to this year are already highly uncertain, any prediction beyond this point would be subject to enormous imponderables with regard to both the future dominant energy technologies (and the resulting emissions) and the variability of the solar dynamo.

\subsection{Predicting CO$_2$}
Let us begin by discussing  the most plausible evolution of the
atmospheric CO$_2$ concentration $C(t)$ until 2100. 
This requires  estimates of the time-dependent sources (emissions) $E(t)$ and the sinks $S(t)$. Quite generally, 
$C(t)$ is governed by the rate equation:
\begin{equation}
\frac{d C(t)}{d t}= E(t)+S(t)   \;.
\end{equation}
Regarding the source term $E(t)$, I will rely solely on the International Energy Agency's (IEA) recent forecasts until 2050 \cite{IEA2025}, rather than considering the wide variety of representative concentration pathways (RCPs) or shared socioeconomic pathways (SSPs).
The most reasonable estimates for the current and stated policy scenarios until 2050 were shown in Figures 3.3 and 4.3 
of \cite{IEA2025}. Remarkably, under the current policies, the emission is expected to remain almost unchanged at around 38\,Gt/yr. In a more ``optimistic'' scenario, the implementation of the stated policies would result in a decrease to 30\,Gt/yr.
Considering the generally increasing energy demand in parallel with the growing proportion of renewable and nuclear energy, it seems most plausible to assume a relatively constant emission scenario of 40\,Gt/yr even until the year 2100\footnote{Such an IEA-based scenario is at least much more realistic than the so-called ``business as usual'' RCP8.5 scenario, which has been thoroughly debunked \cite{Hausfather2020,Pielke2021}.}.
However, I will also examine values that are slightly 
lower (30\,Gt/yr) or slightly higher (50\,Gt/yr), thereby covering a certain plausible range.

Now let us consider the sink term $S(t)$. It is well known that at present approximately half of the current emissions are absorbed by the oceans and land plants \cite{Dengler2023}, and that this absorption increases as the actual CO$_2$ concentration diverges further from its pre-industrial equilibrium value $C_{\rm eq}=280$\,ppm. However, the details of this concentration dependence are not well known.  Corresponding considerations typically fall under the notion of {\it{carbon-concentration feedback}}.
While these feedbacks were initially discussed \cite{Friedlingstein2006,Zickfeld2011} in terms of {\it time-integrated} flux changes due to an increase in $C(t)$ (and corresponding changes due to a temperature increase, a minor effect that is ignored here), the {\it instantaneous} version of Boer and Arora \cite{Boer2013}  appears more understandable and useful. 
These authors set out from the premise that $S(t)$ can be expanded in a Taylor series in $C(t)-C_{\rm eq}$, i.e. around the equilibrium value $C_{\rm eq}=280$\,ppm, beginning with the linear term.
As in \cite{Boer2013}, here we use the ansatz $S(t) = B(t)(C(t) - C_{\rm eq})$.
Although $B(t)$ can be considered constant for small values of $C(t)-C_{\rm eq}$, it may deviate from this for larger values, which would correspond to terms higher than linear in the Taylor expansion.
According to Figure 4 of \cite{Boer2013} 
this deviation depends, in turn, on the emission scenario.
While, for the
unrealistic scenario RCP8.5, $B(t)$ remains rather constant at appr. -0.04\,GtC/(yr ppm), it 
weakens for the 
RCP4.5 scenario to values of -0.03\,GtC/(yr ppm) 
or even -0.02\,GtC/(yr ppm). Later on, we will consider such modified values of $B(t)$ as alternative scenarios.

For now, however, we will stick to the 
constant term and set it to $B=-0.02$\,yr$^{-1}$. This value was chosen based on recent empirical estimations \cite{Dengler2023,Dengler2024}. Yet, it is also compatible with the above value of -0.04\,GtC/(yr ppm) \cite{Boer2013} when considering the equivalence  1\,ppm $\widehat{=}$ 2.124\,GtC $\widehat{=}$7.782\,GtCO$_2$.

\begin{figure}[H]		
\includegraphics[width=0.7\linewidth]{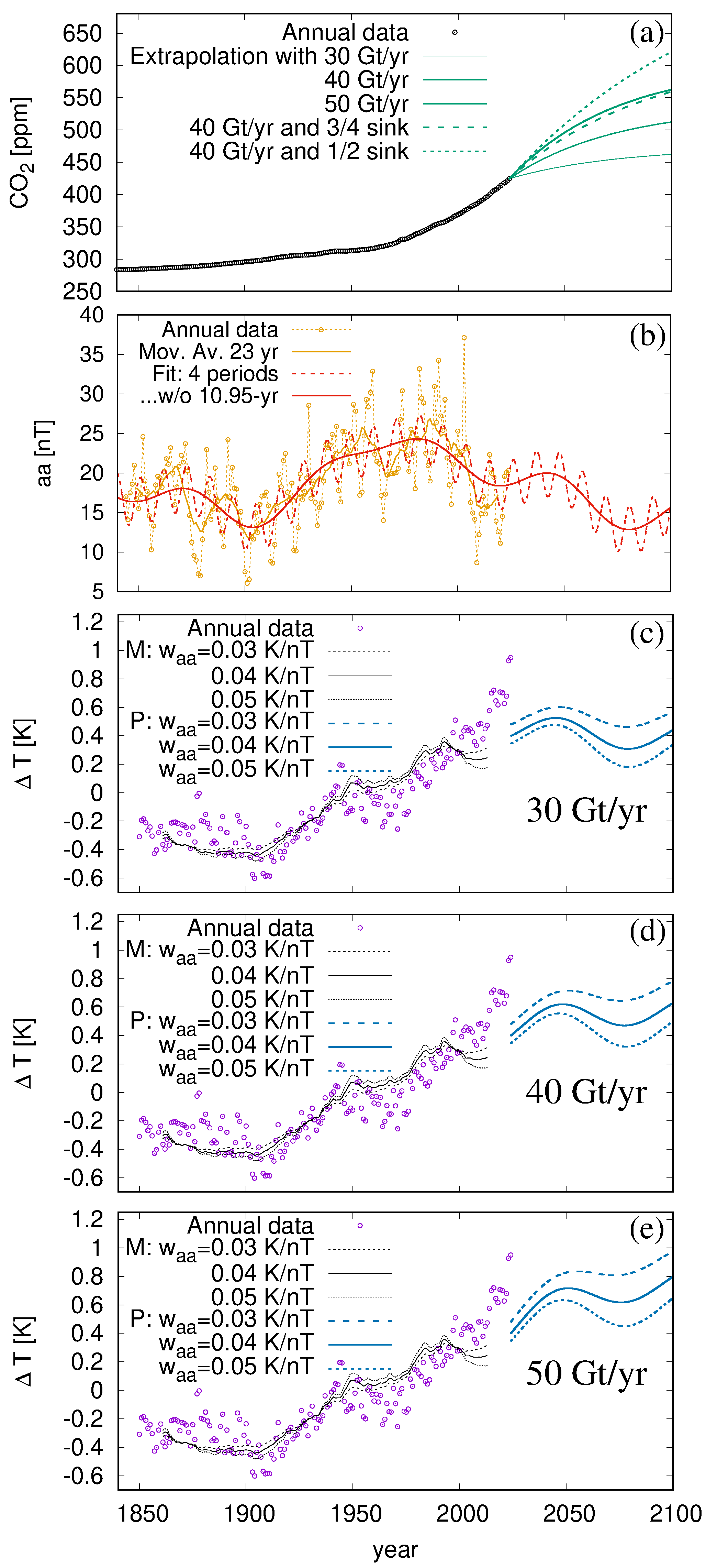}
\caption{Data until 2024, and forecasts until 2100. (a) CO$_2$ content according to Equation (3)  with sink parameter $B=-0.02$\,yr$^{-1}$ and three annual emission values (30, 40, 50\,Gt), and with two reduced sink terms (only for 40\,Gt). (b) Best fit of the aa-index by three dominant periods (and a Schwabe-type period of 10.95\,years), according to the synchronization model of \cite{Stefani2024}, and its extrapolation until 2100. (c) Temperature data, their modelings (M) according to Figure 6b,c,d, and predictions (P) for three 
parameter combinations of $w_{\rm aa}$ (0.03\,K/nT, 0.04\,K/nT, 0.05\,K/nT) and $w_{\rm{CO_2}}$ (1.43\,K,
1.26\,K and 1.09\,K). For the predictions, annual emissions of 30\,Gt are assumed. (d) Same as (c), but for   40\,Gt.  (e) Same as (c), but for 50\,Gt.} 
\label{FIG:fig7}
\end{figure}

Assuming  a constant emission term $E$ and a 
constant value of $B$, the resulting simple
differential equation 
\begin{equation}
\frac{d C(t)}{d t}= E+B (C(t)-C_{\rm eq})   
\end{equation}
acquires the  analytical solution
\begin{eqnarray}
C(t)= K \exp{(B(t-t_0))}-(E/B-C_{\rm eq}) \; ,
\end{eqnarray}
with the integration constant 
$K=C(t_0)+(E/B-C_{\rm eq})$. In the following,
$t_0$ will be set  to 2024, with
$C(t_0=2024)=424.6$\,ppm.

In Figure 7a, the solid green lines show the solutions when setting, in Equation (3), $E$ to either 30\,Gt/yr, 40\,Gt/yr or 50\,Gt/yr, always choosing $B=0.02$\,yr$^{-1}$.
As we can see, in the year 2100 the curves terminate at values between 460\,ppm and 560\,ppm, nearly approaching the respective equilibrium values 473\,ppm, 537\,ppm and 601\,ppm that would result (when setting the r.h.s. of Equation (2) to zero) for annual emissions of 30, 40, or 50\,Gt, respectively. Only for the case of 40\,Gt, two further curves 
are added corresponding to the absorption
coefficient $B$ being reduced by a factor 3/4 
(dashed line)  or
1/2 (dotted line), as inspired by Figure 4 of \cite{Boer2013}. As can be seen, the curve 
for 40\,Gt with $B$ reduced to 3/4 
is very close to the curve for 50\,Gt/yr
with the full $B$. Hence, later on we will focus  
only on the three different emission
scenarios, keeping $B$ always fixed at $0.02$\,yr$^{-1}$.

\subsection{Predicting the solar dynamo}

Having explained our choice of the most plausible CO$_2$ scenarios until 2100, I now turn to forecasting the aa-index.
It is well known that the aa-index (as well as the  sunspot number, with which aa is 
strongly correlated \citep{Cliver1998a}), is governed by the solar dynamo. Clearly, any forecast of the aa-index hinges on predicting the solar dynamo.

As in \cite{Stefani2021}, the aa-forecast is based on the synchronised solar dynamo model, which we have corroborated over the last decade
\citep{Weber2015,Stefani2016,Stefani2018,Stefani2019,Stefani2020a,Stefani2020b,Stefani2021b,Stefani2023,Klevs2023,Horstmann2023,Stefani2024,Stefani2025}. While this model had been the subject of controversy until recently \cite{Nataf2022,Weisshaar2023}, its latest embodiment \citep{Horstmann2023,Stefani2024,Stefani2025}, which is based on the tidal excitation of magneto-Rossby waves at the solar tachocline, shows remarkable agreement with observations.  Firstly, the model provides an extremely accurate representation of the period of the quasi-biennial oscillation (QBO), viz., 1.723\,years in the model versus 1.724 years based on observations of ground-level enhancement events \cite{Velasco2018,Stefani2025}.
Secondly, when also incorporating the spin-orbit coupling effect due to the rosette-shaped motion of the Sun around the solar system's barycenter, the model yields dominant spectral peaks that are in stunning agreement with those of sediment data from Lake Lisan (see Figure 9 in  \cite{Stefani2024})\footnote{
The latter agreement also points to a significant top-down effect whereby stratospheric-tropospheric coupling influences the trajectories of North Atlantic cyclones \cite{Veretenenko2023} which presumably are responsible for the  precipitation rates in the Lake Lisan region \cite{Prasad2004}.}.

Given the model's ability to explain the three dominant periods, it is recommended that the resulting values for the Suess-de Vries cycle (193\, years) and the two Gleissberg-type cycles (57 and 90\,years) be employed for fitting the aa-index data.
To mitigate any end effects, a Schwabe cycle was added as a fourth period. Instead of the theoretical value of 11.07\,years resulting from our model \cite{Stefani2024}, the slightly modified value of 10.95\,years is used, according to the empirical analysis of \cite{Takalo2021}.

Figure 7b shows the resulting fit of the aa-index data, represented by the dashed red line.
The long-term part (full red line), with the 10.95\,yr contribution subtracted, is then extrapolated until 2100. Evidently, the 
interplay of one Suess-de Vries and two Gleissberg cycles predicts 
a shallow maximum around 2045 and a 
minimum around 2080 which is as deep as
the so-called ``Gleissberg minimum''
around 1900 \cite{FeynmanRuzmaikin2014}.

\subsection{Predicting the temperature}

Using these CO$_2$ and aa-index predictions, I will now 
attempt to forecast the global mean temperature until 2100.
To achieve this, I first assume (full blue line in Figure 7d) 
a conservative emission scenario with 40\,Gt/yr, 
an aa-sensitivity of 0.04\,K/nT  and a corresponding 
CO$_2$-sensitivity of 1.26\,K. While this combination 
is somehow similar to the middle scenario shown 
as the green line in Figure 9d of \cite{Stefani2021}, its 
higher aa-sensitivity leads to a more oscillatory behavior 
of the temperature curve.
Following a slight increase until 2050 and a modest 
decline until 2080, the temperature reaches 0.6\,K in 2100.
The dashed and dotted blue lines added in Figure 7d refer to 
decreased (0.03\,K/nT) and increased (0.05\,K/nT) values 
of $w_{\rm aa}$ of 0.03 K/n, combined with the respective values of 
1.4\,K and 1.1\,K for $w_{\rm CO_2}$.

Figures 7c and 7e show the corresponding
curves for (hypothetical) reductions or enhancements 
of the annual emissions to 30\,Gt or
50\,Gt, respectively. Roughly speaking, these
emission changes lead to an decrease or increase in the 
final temperature of around 0.2\,K.

\section{Summary and discussion}

The aim of this work was to update and refine the previous estimate \cite{Stefani2021} of the influence of solar activity variations and CO$_2$ on the terrestrial climate, and to improve the global temperature prediction until the end of this century.
Having supplemented the underlying dataset with information from the last six years (2019–2024), an attempt was made to narrow the broad range of climate sensitivities (on both the aa-index and CO$_2$ doubling) that had resulted from the double regression in \cite{Stefani2021}.
For two different lengths of the moving average window I have compared the time-dependence of the goodness of fit for the ``statistically correct'' double regression with 
those for single regressions on aa or CO$_2$, respectively.
It was shown that, until the end of the 20$^{\rm th}$ century, the $R^2$ value of the double regression was almost identical to that obtained using the aa-index alone. This indicates that the Sun's impact on climate was the dominant factor up to that point\footnote{This result not only contrasts with the original idea of \cite{Callendar1938}  and the subsequent claim of \cite{Mann1998}  that  ``(t)he partial
correlation with CO$_2$ indeed dominates over 
that of solar irradiance for the most recent 200-year interval'', but it also contradicts 
the more Solomon-like judgment of  \cite{Solanki2003} that 
``since roughly 1970 the solar influence on climate...cannot have been dominant''.}. 
Moreover, when assessing the {\it adjusted} $\overline{R}^2$ until 1985 (using an MAW of 11 years) or 2005 (using an MAW of 23 years), the single regression on aa alone yields higher values than the double regression.
It was only since the beginning of the 21$^{\rm st}$ century that the contribution of CO$_2$ has become 
noticeable and prominent.

Given the remarkable stability of the aa-sensitivity resulting from the double and single regressions over the long period between 1950 and 1990, the outcome of around 0.04 K/nT was considered plausible for subsequent decades as well.
Based on this reasoning, the CO$_2$ sensitivity was recalculated by first subtracting the aa-contribution (weighted by the 
aa-sensitivity obtained beforehand) from the observed $\Delta T$ and then regressing the remaining $\Delta T$ on CO$_2$. 
Although this procedure may appear to be
a ``statistical sin'', I consider its outcome to be better justified
than that of a ``correct'', albeit somewhat naive double-regression.
That way, 
the CO$_2$-sensitivity was narrowed down from the
ample range of 0.6-1.6\,K derived in \cite{Stefani2021} to  1.1-1.4\,K.
The alternative double regression 
would not only have resulted 
in a moderately enhanced CO$_2$-sensitivity 
value of 1.74\,K, but also in
a dramatically reduced aa-sensitivity of
0.011\,K/nT.  Faced with the choice of either accepting an unreasonably huge reduction in the aa-sensitivity over the last two decades (compared with its stable long-term average) or attributing some of the recent steep warming to factors other than CO$_2$ or the Sun, I opted for the latter.
While the recent temperature decline until the end of 2025  bodes well for the correctness of our approach only future data will give a definite answer.

In the parlance of global circulation models, the derived CO$_2$ sensitivity should be considered a type of TCR rather than ECS. The obtained range of 1.1-1.4\,K lies at the lower end of the values published by the IPCC (1.2–2.4\,K) \cite{AR6}, but corresponds reasonably with those 
of Lewis and Curry \cite{Lewis2018} and Scafetta \cite{Scafetta2023a}. As the values of $\Delta T$ in the final years of the period under consideration were particularly high, it seems likely that including data from the following years will result in a slight decrease in our values. Although it is highly unlikely, we cannot rule out the possibility that the recent rise in temperature is largely due to CO$_2$. In this case, however, the aa-sensitivity would have had to have decreased significantly in recent decades for some unknown reasons.

 In the second part of the paper, I have made another attempt to forecast the global temperature until the end of the century based on the sharpened ranges of the two climate sensitivities. 
In addition to that reduction, I also limited the potential CO$_2$ history to three variants. All of them rely on constant annual emissions (30, 40 or 50 Gt) over the next few decades (see \cite{IEA2025}), and on a simple linear sink model, which I consider most plausible in view of the results of \cite{Dengler2023,Dengler2024}.

As for the prediction of the aa-index, the variety of models of 
\cite{Stefani2021} was restricted 
to a single one 
comprising the three dominant
periods (one Suess-de Vries and two Gleissberg cycles) according to our synchronized solar dynamo model. 
Although this may seem overly specific, the use of this model was encouraged by the fact that its resulting spectrum corresponds remarkably well with climate-related sediment data from Lake Lisan \cite{Stefani2024,Prasad2004}.
 
The combined forecast model yields, for the terminal year 2100, a rather benign temperature increase of appr. 0.6\,K, compared to the standard HadSST data reference period (1961–1990). For each emission scenario, the temperature spread between the considered aa-sensitivity models was approximately 0.3\,K, whereas the corresponding spread between different emission scenarios was around 0.4\,K.

Only for the most ``pessimistic'' 
parameter combination (assuming high 
annual emissions of 50\,Gt, a low aa-sensitivity of 
0.03\,K/nT, and an accordingly
high CO$_2$ sensitivity of 1.43\,K) 
I obtained a temperature rise to slightly 
more than 1\,K. This essentially corresponds to the extraordinarily high value measured in 2024. For all other parameter combinations we would stay below this value.  
When compared with the minimum temperature in the dataset (appr. -0.4\,K around 1900), the increase is no greater than 1.5 K. Regardless of the actual relevance of the politically pursued 1.5-degree target, there seems to be no danger of exceeding it if we assume that emissions will remain at current levels.

\section{Conclusions and outlook}

All forecast were deliberately restricted until 2100, primarily because of the significant uncertainty surrounding the CO$_2$ emission scenarios.  
While neither a huge increase in emission values (as 
in the former scenario A2 \cite{Friedlingstein2006} 
or the recent scenario RCP8.5 \cite{Boer2013}), nor a dramatic decrease to an illusory ``net zero'' regime appears very likely, it would be preposterous to prophesy the energy production landscape in one hundred years' time.

A similar caveat applies to solar activity. While the dominant periods used here are typical of a regular solar dynamo behaviour, we cannot rule out the occurrence of a new grand solar minimum (i.e. the next Bond event) in the centuries to come. Transitions between regular and chaotic regimes of the solar dynamo, termed supermodulation in \cite{Weiss2016}, appear to be non-deterministic \cite{Stefani2021b}, although the Bray-Hallstadt cycle (appr. 2300\,years) may also play a certain role here \cite{Beer2018}.

Finally, we should always bear in mind that we are amidst the decline phase of the glacial cycle which is governed by changing insolation conditions, dominated by the Milankovic cycles. 
Interestingly, glaciation always occurred when the nutation angle of the Earth decreased below 23$^{\circ}$, though
deglaciation was sometimes missed when it increased above 23$^{\circ}$. This effect was recently discussed by Vin\'{o}s \cite{Vinos2022}, who also employed it to solve the longstanding 100,000-year problem.

In any case, it is only when considering the ECS, and its century-long relaxation mechanism, that those long-term effects might acquire any relevance. 
Even assuming yearly CO$_2$ emissions of 30, 40 or 50 Gt, the predicted benign temperature rise until 2100 gives us ample time to develop cost-efficient alternatives to fossil fuels that will not result in an unnecessary reduction in living standards\footnote{
Note that 40\,Gt/yr divided by the present world population of 8 billion equates to the 
5\,t/yr of specific French emissions. Even without an increase in current global emissions, it would hence be possible for everyone to live like ``God in France'' (admittedly a country that generates a high proportion of its electricity from nuclear power).}.

\funding{This work received
funding from the Helmholtz Association in frame of the AI project GEOMAGFOR (ZT-I-PF-5-200).}

\dataavailability{
The see surface temperature data HadSST.4.2.0.0 data 
were obtained from 
http://www.metoffice.gov.uk/hadobs/hadsst4 on November 27, 
2025 and are $\copyright$ of
British Crown Copyright, Met Office (2020).
The aa-index data between 1868 and 2010 were obtained from
NOAA under
ftp://ftp.ngdc.noaa.gov/STP/GEOMAGNETIC$\_$DATA/AASTAR/.
Monthly aa-index data from 2011-2024 were obtained
from the webpage of the British Geological Survey, 
www.geomag.bgs.ac.uk/data$\_$service/data/magnetic$\_$indices/aaindex.html.
CO$_2$ date were obtained from www.co2.earth,  www.co2.earth  and gml.noaa.gov.}

\acknowledgments{I would like to express my gratitude to Joachim Dengler, Tom Weier and Willie Soon for valuable feedback on an early draft of this paper.}

\conflictsofinterest{The author declares no conflict of interest.}

\end{paracol}
\reftitle{References}

\end{document}